\documentclass[11pt]{article}

\usepackage[T1]{fontenc}
\usepackage[utf8]{inputenc}
\usepackage{amsmath,amssymb,amsthm}
\usepackage{enumerate}
\usepackage{color}
\usepackage{mathtools}
\usepackage{thm-restate}
\usepackage{tikz}

\usepackage[Algorithm,ruled]{algorithm}
\usepackage[noend]{algpseudocode}
\usepackage{cite}

\usepackage{geometry}
\definecolor{darkgreen}{rgb}{0,0.5,0}
\usepackage{hyperref}
\hypersetup{
    unicode=false,          
    colorlinks=true,        
    linkcolor=blue,          
    citecolor=darkgreen,        
    filecolor=magenta,      
    urlcolor=cyan           
}
\usepackage[capitalize, nameinlink]{cleveref}
\crefname{theorem}{Theorem}{Theorems}
\crefname{lemma}{Lemma}{Lemmas}
\crefname{corollary}{Corollary}{Corollaries}
\crefname{observation}{Observation}{Observations}

\renewcommand{\epsilon}{\varepsilon}

\newtheorem{theorem}{Theorem}[section]
\newtheorem{lemma}[theorem]{Lemma}

\newtheorem{definition}[theorem]{Definition}

\newtheorem{observation}[theorem]{Observation}

\newcommand{\set}[1]{\left\{#1\right\}}
\newcommand{\nca}{\ensuremath{\mathrm{nca}}}

\title{Nearest Common Ancestors:\\
Universal Trees and Improved Labeling Schemes}

\author{ 
Fabian Kuhn%
\thanks{Department of Computer Science, University of Freiburg, Germany.}
\and
Konstantinos Panagiotou%
\thanks{Department for Mathematics, University of Munich, Germany}%
\and
Pascal Su%
\thanks{Department of Computer Science, ETH Zurich, Switzerland. }%
}

\date{}

\begin{document}
\maketitle

\begin{abstract}
  We investigate the nearest common ancestor (NCA) function in rooted trees. As the main conceptual contribution, the paper introduces universal trees for the NCA function: For a given family of rooted trees, an NCA-universal tree $S$ is a rooted tree such that any tree $T$ of the family can be embedded into $S$ such that the embedding of the NCA in $T$ of two nodes of $T$ is equal to the NCA in $S$ of the embeddings of the two nodes.

  As the main technical result we give explicit constructions of NCA-universal trees of size $n^{2.318}$ for the family of rooted $n$-vertex trees and of size $n^{1.894}$ for the family of rooted binary $n$-vertex trees. A direct consequence is the explicit construction of NCA-labeling schemes with labels of size $2.318\log_2 n$ and $1.894\log_2 n$ for the two families of rooted trees. This improves on the best known such labeling schemes established by Alstrup, Halvorsen and Larsen [SODA 2014].
\end{abstract}


\section{Introduction}

The nearest common ancestor\footnote{In the literature, the nearest common ancestor of two vertices in a rooted tree is sometimes also referred to as the lowest or least common ancestor (LCA).} (NCA) of two vertices $u$ and $v$ of a rooted tree $T$ is the first common vertex of the paths connecting $u$ and $v$ to the root of $T$. Finding the nearest common ancestor appears as an essential operation in many algorithms and applications (see for example the survey by Alstrup, Gavoille, Kaplan, and Rauhe \cite{Alstrup:2002:NCA:564870.564914}).

\paragraph{NCA-Universal Trees.} The present paper introduces the notion of \emph{NCA-universal trees} as a novel tool to study and algorithmically deal with the NCA function in rooted trees. We define an NCA-universal tree $S$ for a family of rooted trees $\mathcal{T}$ as such that every tree $T\in\mathcal{T}$ can be embedded into $S$ such that the NCA function is preserved by the embedding. More formally, an embedding of $T$ into $S$ is an injective mapping $\varphi_T$ of $V(T)$ into $V(S)$ such that the embedding function $\varphi_T$ and the NCA function commute.

\paragraph{NCA-Labeling Schemes.} As an immediate application of an NCA-universal tree $S$ for a family $\mathcal{T}$ of rooted trees, $S$ directly implies an \emph{NCA-labeling scheme} \cite{Alstrup:2002:NCA:564870.564914} for the family $\mathcal{T}$. Generally, a labeling scheme is a way to preprocess the structure of a graph to later allow simple and fast queries. A labeling scheme consists of an encoder and a decoder, where the encoder must be able to label a family of graphs such that the decoder can answer queries, given just the labels and no additional information about the underlying graph. More specifically, an NCA-labeling scheme assigns a unique label to each node of a rooted tree $T$ such that given the labels of two vertices $u$ and $v$ of $T$, it is possible to compute the label of the NCA of $u$ and $v$ in $T$. If an NCA-univeral tree $S$ for a family $\mathcal{T}$ of rooted trees is given, we can get an NCA-labeling scheme for $\mathcal{T}$ as follows. Let $|S|$ be the number of vertices of $S$ and assume that the vertices of $S$ are labeled from $0$ to $|S|-1$ in a arbitrary fixed way. Given an embedding of a tree $T\in\mathcal{T}$ into $S$, we then get the labeling of a vertex $v$ of $T$ by using the label of the vertex $x$ of $S$ to which $v$ is embedded. The size of the labels (in bits) of the labeling scheme is therefore exactly $\lceil\log|S|\rceil$. We remark that throughout the paper, all logarithms are to base $2$.

\paragraph{Contribution.} We show that the family of all rooted trees with at most $n$ vertices has an NCA-universal tree of size $O(n^{2.318})$ and that the family of all binary rooted trees with at most $n$ vertices has an NCA-univeral tree of size $O(n^{1.894})$. This implies that the families of rooted $n$-vertex trees and of rooted $n$-vertex binary trees have labeling schemes with labels of size $2.318\log n$ and $1.894\log n$, respectively. This improves on the best previous NCA-labeling schemes that were developed by Alstrup, Halvorsen and Larsen \cite{Alstrup:2014:NLS:2634074.2634146} and which require labels of size $2.772\log n$ for general rooted trees and of size $2.585\log n$ for binary rooted trees. In \cite{Alstrup:2014:NLS:2634074.2634146}, it is also shown that any NCA-labeling scheme for general $n$-vertex rooted trees requires labels of size at least $1.008\log n$.

As we show how to explicitly construct the NCA-universal trees, our labeling schemes are constructive. Note that the best NCA-labeling schemes of \cite{Alstrup:2014:NLS:2634074.2634146} are not constructive and that the best previous constructive NCA-labeling scheme for $n$-vertex rooted trees requires labels of size $3\log n$. Further, our NCA-labeling schemes are \emph{efficient}, the embedding of a rooted tree into the constructed NCA-universal tree can be computed efficiently and a single query can be answered in time $O(\log^2n)$ ($O(\log n)$ for binary trees). We believe that our new NCA-labeling schemes are not only interesting because they improve upon the best existing schemes, but also because our approach leads to more intuitive and significantly simpler constructions.

\paragraph{Related Work.} Graph labeling schemes are an elegant way
to store structural information about a graph. As every vertex is only
assigned a small label, the information is stored in a completely
distributed way and graph labelings therefore are particularly
interesting in a distributed context, where labeling schemes are used for various kinds of graph queries \cite{Gavoille2003,Peleg1999}. In addition, labeling schemes can be used in a context where extremely large graphs are processed and where accessing the data is expensive. To answer a pair-wise query, only the two labels of the corresponding vertices need to be accessed.

The first labeling schemes that appear in the literature are adjacency labeling schemes (given the labels of two vertices, determine whether the vertices are adjacent). They were introduced among others by Breuer \cite{1053860} and Folkman \cite{BREUER1967583}. In the context of adjacency labeling schemes, it is well known that they are tightly connected to induced universal graphs. A graph $G$ is an induced universal graph for a graph family $\mathcal{H}$ if $G$ contains every graph $H\in \mathcal{H}$ as an induced subgraph. Induced universal graphs were first described by Rado \cite{Rado1964} and Kannan et al.\cite{doi:10.1137/0405049} noted the equivalence between adjacency labeling schemes and induced universal graphs.

When considering the family of rooted trees we are interested in  different queries such as whether a vertex is an ancestor of the other. Ancestry labeling has been studied and labeling schemes of size $\log(n) + \Theta(\log(\log(n)))$ are known to be tight \cite{Abiteboul:2001:CLS:365411.365529}. If the tree has low depth, then a scheme of size $\log(n) +2\log(d) +O(1)$ is known, where $d$ is the depth of the tree \cite{Fraigniaud:2010:CAL:1873601.1873639}.

For NCA-labeling schemes, a linear-size labeling scheme that answers queries in constant time was introduced by Harel and Tarjan in \cite{harel1980linear, hareldoi:10.1137/0213024}. In the following, there was a series of significant improvements in \cite{powell1990further}, \cite{bender2000lca}, \cite{Gabow:1984:SRT:800057.808675}, \cite{doi:10.1137/0217079}, \cite{berkman1993recursive}, \cite{Alstrup:2002:NCA:564870.564914} and most recently in \cite{Alstrup:2014:NLS:2634074.2634146}. In particular in \cite{Alstrup:2014:NLS:2634074.2634146}, a lower bound for NCA labeling schemes of $1.008\,\log n$ is shown, which separates NCA-labelings, that need labels of size $\log n + \Omega(\log n)$, from ancestry labeling schemes, where labels of size $\log n + O(\log\log n)$ are sufficient. 

There is concurrent work by Gawrychowski and Łopuszański \cite{1707.06011} who reach the exact same bounds for labeling schemes and construct almost identical universal trees as we do. The proof method is a bit different and in their paper they also have lowerbounds for the size of a universal tree for the NCA function.

\paragraph{Outline.}
The rest of the paper is organized as follows. In the remainder of this section, we first formally define the problems and state our results in \Cref{sec:def,sec:results}. In \Cref{sec:simpleuniversal}, we first prove a simpler upper bound of $O(n^2)$ on the size of NCA-universal trees for the family of binary rooted trees. We extend the construction of \Cref{sec:simpleuniversal} to obtain the stronger and more general results stated above in \Cref{sec:universal}. Finally, in \Cref{sec:impl}, we sketch how to efficiently implement our labeling scheme and in \Cref{sec:concl}, we conclude the paper and discuss some open issues.

\subsection{Definitions}
\label{sec:def}

We next define the necessary graph-theoretic concepts and notation and we in particular formally introduce the notion of universal trees for the NCA function. In the following let $\mathcal{T}_n$ be the family of unlabeled rooted trees with at $n$ vertices and $\mathcal{B}_n$ the family of unlabeled rooted binary trees with $n$ vertices. 

\begin{definition}\label{def:ancestor}
 In a rooted tree $T=(V,E)$ a vertex $u$ is an \textbf{ancestor} of a vertex $v$ if $u$ is contained in the (unique) path from $v$ to the root of $T$.  
\end{definition}

Note that according to the above definition, a node $u$ is an ancestor of itself.

\begin{definition}\label{def:nca}
Let $T=(V,E)$ be a rooted tree. For a pair of vertices $u$ and $v$ their \textbf{nearest common ancestor (NCA)} $\nca_T(u,v)$ is the unique common ancestor that is furthest from the root of $T$. 
\end{definition}
With this notation at hand we can define the notion of a universal tree for NCA.
\begin{definition}\label{def:universaltree}
A rooted tree $S$ is called an \textbf{NCA-universal tree} for a family of rooted trees $\mathcal{T}$, if for every tree $T\in \mathcal{T}$ there is an embedding function $\varphi_T:V(T)\mapsto V(S)$ such that $\varphi_T$ commutes with the NCA function, i.e., for all $u,v\in V(T)$, $\varphi_T(\nca_T(u,v)) =\nca_S(\varphi_T(u),\varphi_T(v))$.
\end{definition}

Hence, the embedding has the property that the NCA of two nodes $u$ and $v$ of $T$ is mapped to the NCA of $\varphi_T(u)$ and $\varphi_T(v)$ in $S$. Note that we do not require the root of $T$ to be embedded to the root of $S$. In the following, a rooted tree that is universal for the NCA function is also called an \emph{NCA-universal tree}. 

\begin{definition}
An \textbf{NCA-labeling scheme} for a family of rooted trees $\mathcal{T}$ is a pair of functions called the encoder ($f$) and decoder ($g$) with $ f : \{ v|v\in T \in \mathcal{T} \} \mapsto [m] $ and $g:[m]\times [m] \mapsto [m]$ satisfying the following properties.
\begin{enumerate}[i)]
  \item for every $T\in \mathcal{T}$ and every $u,v\in V(T)$, $f(u)\neq f(v)$ and
  \item for every $T\in \mathcal{T}$ and every $u,v\in V(T)$, $g(f(u),f(v))=f(\nca_T(u,v))$.
\end{enumerate}
For a node $v\in T\in \mathcal{T}$, $f(v)$ is called the label of $v$. The size of the labeling scheme defined by $f$ and $g$ is $\lceil\log m\rceil$, i.e., the number of bits required to store the largest label.
\end{definition}

Given an NCA-universal tree $S$ for a family of rooted trees $\mathcal{T}$, we directly obtain an NCA-labeling scheme for $\mathcal{T}$.

\begin{observation}
\label{obs:unilabel}
  Let $S$ be an $N$-vertex rooted tree that is universal for the NCA function and the family $\mathcal{T}$ of rooted trees. Then, there exists an NCA-labeling scheme of size $\lceil\log N\rceil$ for $\mathcal{T}$.
\end{observation}
\begin{proof}
  We assign unique names from $0$ to $N-1$ to the $N$ vertices of $S$. Consider a tree $T\in \mathcal{T}$ and let $f$ be an embedding of $T$ into $S$. The label of a vertex $v$ of $T$ is the name assigned to vertex $f(v)$ of $S$. Given the labels $x_u\in\set{0,\dots,N-1}$ and $x_v\in\set{0,\dots,N-1}$ of two nodes $u$ and $v$ of $T$, the decoder outputs the name $x_w\in\set{0,\dots,N-1}$ of the NCA of the vertices $u'$ and $v'$ with names $x_u$ and $x_v$ in $S$.
\end{proof}

\subsection{Main Results}
\label{sec:results}
Our main result is an explicit construction of universal trees for the families of all trees and binary trees with $n$ vertices. The same bounds can be found in the concurrent work of Gawrychowski and Łopuszański \cite{1707.06011}.
\begin{theorem}
\label{thm:universal}
Let $n\in\mathbb{N}$. Then:
\begin{itemize}
	\item There is a rooted tree $S_n$ of size less than $n^{ 2.318}$ that is universal for the NCA function and the set $\mathcal{T}_n$ of rooted trees of size $n$.
	\item There is a rooted tree $S_{n}^{bin}$ of size less than $n^{1.894}$ which is universal for the NCA function and the set $\mathcal{B}_n$ of rooted binary trees of size $n$.
\end{itemize}
\end{theorem}
The proof can be found in Section 3. The direct implication of this for labeling schemes is summarized in the following statement.
\begin{theorem}
\label{thm:scheme}
For any $n\in \mathbb{N}$ there exists an $NCA$-labeling of size less than $ 2.318 \, \log n$ and an $NCA$-labeling for binary trees of size less than $1.894 \,\log n$.
\end{theorem}
This is an improvement of the current best known bound of $2.772 \,\log n$ from \cite{Alstrup:2014:NLS:2634074.2634146}. Further for the specific case of binary trees of particular interest is that the constant is now below $2$ and therefore will likely not be an integer. 
\begin{proof}[{\rm\bf Proof of \Cref{thm:scheme}}] This is exactly what we have shown in \Cref{obs:unilabel} and therefore follows from \Cref{thm:universal}.
Although this is just an existential proof, from the construction we will see later, it is clear that a reasonably fast algorithmic implementation is possible and we will give a sketch in \Cref{sec:impl}.
\end{proof}

\section{Basic Universal Tree Construction}
\label{sec:simpleuniversal}

The NCA-universal trees of \Cref{thm:universal} are constructed recursively. 
Before proving the general statements of \Cref{thm:universal}, we describe a simpler, slightly weaker construction that provides an NCA-universal tree of size $O(n^2)$ for $n$-vertex binary trees. The full constructions required to prove \Cref{thm:universal} appears in \Cref{sec:universal}.

\begin{theorem}\label{thm:reduced}
  For any $n\in \mathbb{N}$ there exists a rooted tree $S_n$ of size
  less than $n^{2}$ which is universal for the NCA function and the
  rooted binary trees of size at most $n$.
\end{theorem}

Our recursive universal tree construction requires two kinds of NCA-universal trees. In addition to ordinary unlabeled rooted trees, we also need to define NCA-universal trees for the family of rooted binary trees where one leaf node is distinct (marked). Recall that $\mathcal{B}_n$ denotes the set of all $n$-vertex unlabeled rooted binary trees, so let $\mathcal{B}_n'$ denote the family of unlabeled rooted binary trees on at most $n$ vertices and with one marked leaf.

\begin{definition}\label{def:universaltreemarked}
A rooted tree $S'$ with one marked leaf vertex $w$ is called a \textbf{NCA-universal tree} for a family of rooted trees $\mathcal{T}'$ with \textbf{one marked leaf} if for every tree $T\in \mathcal{T}'$, there exists an embedding function $\varphi_T : V(T) \mapsto V(S')$ that maps the marked leaf of $T$ to the marked leaf $w$ of $S'$ and where $\varphi_T$ commutes with the NCA function, i.e.,  $\varphi_T(\nca_T(u,v)) =\nca_S(\varphi_T(v),\varphi_T(u))$ for all $u,v\in T$.
\end{definition}
As in \Cref{def:universaltree}, we do not require that the root of $T$ is mapped to the root of $S'$.

\begin{figure}[t]
  \centering
  \includegraphics[width=0.8\textwidth]{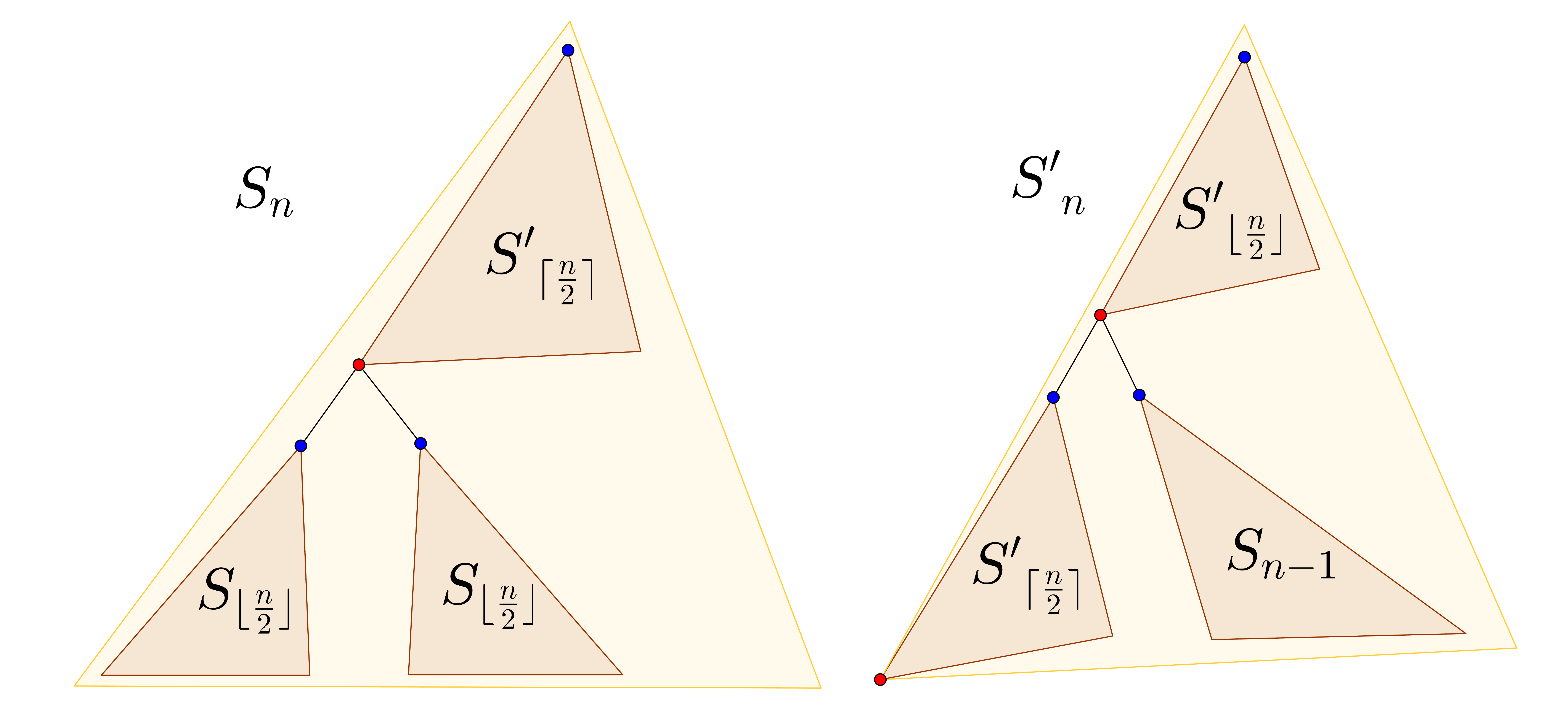}
  \caption{Recursive structure of the basic NCA-universal tree construction for rooted binary trees of size at most $n$.}
  \label{fig:basicrecursion}
\end{figure}

\paragraph{Overview of the Construction.} The construction of the NCA-universal tree for binary rooted trees is done recursively as illustrated in \Cref{fig:basicrecursion}. The universal tree $S_n$ for binary trees of size at most $n$ consists of three NCA-universal trees for binary trees of size at most $n/2$, where one of these three universal trees needs to work for the more general family trees with one marked leaf vertex. Universal trees for the family of $n$-vertex trees with a marked leaf are constructed recursively in a similar way. They consist of two NCA-universal trees for $n/2$-vertex binary trees with a marked leaf and of a single NCA-univeral tree for ordinary $n-1$-vertex binary trees (see \Cref{fig:basicrecursion}).

In order to show that the recursive construction of \Cref{fig:basicrecursion} results in an NCA-universal tree we need to argue that any $n$-vertex binary tree $T$ can be embedded. To achieve this, we show that any rooted tree $T$ has a vertex $v$ such that $v$ splits $T$ into three subtrees of size at most $n/2$. Vertex $v$ is then embedded to the vertex marked in red in the left part of \Cref{fig:basicrecursion}. The three subtrees of $T$ induced by $v$ are then embedded recursively into the three parts of the universal tree construction. Further, we need to show that any rooted tree $T$ with a marked leaf can be partitioned in a similar way to be consistent with the recursive structure in the right part of \Cref{fig:basicrecursion}. In the following, we first give the basic technical lemmas required to partition $n$-vertex trees $T$ into the required smaller subtrees. Based on these partitioning results, we the analyze the recursive NCA-universal tree construction in more detail and prove \Cref{thm:reduced}. As the same partitioning lemmas will also be needed in the general NCA-universal tree constructions in \Cref{sec:universal}, they are stated more generally than what we require for the simple construction of the present section.

\begin{lemma}
  \label{lemsplit}
  For every rooted $n$-vertex tree $T$ and for every
  parameter $\lambda \in (0,1]$, there exists a vertex $v\in V(T)$
  such that removing the edges from $v$ to its children splits the
  tree into components such that each component rooted at a child of
  $v$ has size at most $\lfloor (1 - \lambda) \cdot n \rfloor$ and
  such the remaining component containing $v$ and the root of $T$ has
  size at most $\lceil \lambda \cdot n \rceil$.
\end{lemma}
\begin{proof}
  We determine $v$ using the following simple iterative
  procedure. We initialize $v$ to be the root of $T$. We stop the
  procedure as soon as $v$ satisfies the conditions of the lemma. For
  some vertex $u$ of $T$, let $\mathit{size}(u)$ be the number of
  vertices in the subtree rooted at $u$. If $v$ does not split the
  tree as required, we let $w$ be the child vertex of $v$ that
  maximizes $\mathit{size}(w)$ and we set $v:=w$. Since $v$ goes from
  being the root of $T$ to being a leaf of $T$ during this process,
  the component containing the root goes from being of size $1$ to a
  set of size $n$. We claim that for the last vertex $v$ where the
  connected component of the root is still of size at most
  $\lceil \lambda \cdot n \rceil$, the lemma holds.
  
  Clearly, the component with the root is of size at most
  $\lceil \lambda \cdot n \rceil$ and it thus suffices to show that
  all the subtrees of $v$ are of size at most
  $\lfloor (1-\lambda)n\rfloor$. Assume for contradiction that $v$ has a
  child $w$ such that
  $\mathit{size}(w)\geq 1+ \lfloor (1-\lambda)n\rfloor$. Then,
  removing all subtrees of $w$ from $T$ would result in a component of
  size at most $n-\lfloor (1-\lambda)n\rfloor=\lceil \lambda n\rceil$
  and thus $v$ would not be the last vertex for which the connected
  component of the root is of size at most $\lceil\lambda n\rceil$.
\end{proof}

For trees with a marked leaf we can get a similar tree splitting lemma.

\begin{lemma}
  \label{lemsplitmark}
  Given a rooted $n$-vertex tree $T$ with one marked leaf vertex $w$
  and a parameter $\lambda\in(0,1]$. If $n\ge \frac{1}{1-\lambda}$, there exists a vertex
  $v\in V(T)$ such that when removing the edges connecting $v$ to its
  children, $T$ is split into components satisfying the following
  properties. The component containing the root of $T$ and vertex $v$
  has size at most $\lceil \lambda n\rceil$, the component containing
  the marked leaf $w$ has size at most $\lfloor (1-\lambda)n\rfloor$,
  and all other components have size at most $n-1$.
\end{lemma}

\begin{proof}
  Let $r$ bet the root vertex of $T$. We choose $v$ to be last vertex
  on the path from $r$ to $w$ such that when removing the subtrees of
  $v$, the remaining component has size at most
  $\lceil\lambda n \rceil$. $\lceil\lambda n \rceil<n$, so $v$
  cannot be a leaf and thus $v\neq w$. Let $v'$ be the root of the
  subtree of $v$ containing $w$. To prove the lemma, it suffices to
  show that the subtree rooted at $v'$ has size
  $\mathit{size}(v')\leq\lfloor(1-\lambda)n\rfloor$. For the sake of
  contradiction, assume that
  $\mathit{size}(v')\geq \lfloor(1-\lambda)n\rfloor+1$. In this case,
  the total size of all subtrees of $v'$ is at least
  $\lfloor(1-\lambda)n\rfloor$ and thus removing all subtrees of $v'$
  would leave a component of size at most $\lceil \lambda
  n\rceil$. This contradicts the assumption that $v$ is the last
  vertex on the path from $r$ to $w$ for which this is true.
\end{proof}

\begin{proof}[{\rm\bf Proof of \Cref{thm:reduced}.}]
  For every integer $n\geq 1$, we show how to construct an
  NCA-universal tree $S_n$ for the family $\mathcal{B}_n$ of
  $n$-vertex rooted binary trees and an NCA-universal tree $S_n'$ for
  the family $\mathcal{B}'_n$ of $n$-vertex binary rooted trees with
  one marked leaf. We will prove by induction on $n$ that
  $|S_n|\leq n^2$ and that $|S_n'|\leq 2n^2-1$ for all $n\geq 1$.

  For the induction base, note that $S_1$ and $S_1'$ clearly need to
  only consist of a single vertex and we thus have
  $|S_1|=|S_1'|=1$. Thus, the bounds on $|S_n|$ and $|S_n'|$ hold for
  $n=1$.
  For the induction step, assume that $n\geq 2$ and that
  $|S_k|\leq k^2$ and $|S_k'|\leq 2k^2-1$ for all
  $1\leq k<n$. We build the two NCA-universal trees $S_n$ and $S_n'$ by using smaller NCA-universal trees as given in
  \Cref{fig:basicrecursion}. That is, $S_n$ is composed of one copy of $S_{\lceil n/2\rceil}'$ 
  and two copies of $S_{\lfloor n/2\rfloor}$  and $S_n'$ is composed of one copy of
  $S_{\lceil n/2\rceil}'$, $S_{\lfloor n/2\rfloor}'$, and $S_{n-1}$. We need to show
  that the constructed trees $S_n$ and $S_{n}'$ are in fact
  NCA-universal trees and that they satisfy the required size
  bounds. We first show that the trees are or the right size. Using
  the induction hypothesis, we have
  \begin{eqnarray*}
    |S_n|
    & = &
          \big|S_{\lceil\frac{n}{2}\rceil}'\big| +
          2\cdot\big|S_{\lfloor \frac{n}{2}\rfloor}\big|
    \ \le  \ 2 \cdot
             {\left\lceil\frac{n}{2}\right\rceil}^2-1 + 2\cdot
      {\left\lfloor \frac{n}{2}\right\rfloor}^2
    \ \le \ n^2,\quad\text{and}\\
    |S_n'|
    & =  & \big|S_{\lceil\frac{n}{2}\rceil}'\big| +
           \big|S_{\lfloor\frac{n}{2}\rfloor}'\big| +
           |S_{n-1}|
           \ \le \
           2\cdot {\left\lceil\frac{n}{2}\right\rceil}^2 - 1+
           2\cdot {\left\lfloor \frac{n}{2}\right\rfloor}^2 -1 +
           (n-1)^2
           \ \le \ 2 n^2-1.
  \end{eqnarray*}
  It thus remains to prove that the recursive construction of $S_n$
  and $S_n'$ allows to find a proper embedding for every
  $T \in \mathcal{B}_n$ into $S_n$ and every $T' \in \mathcal{B}_n$
  into $S_n'$, respectively.

  To show this, we use \Cref{lemsplit} and \Cref{lemsplitmark}. We
  first show how to construct an embedding $\varphi_T$ of a binary
  $n$-vertex tree $T\in \mathcal{B}_n$ into $S_n$. For this purpose,
  we apply \Cref{lemsplit} with parameter $\lambda=1/2$ to tree
  $T$. Let $v\in V(T)$ be the vertex of $T$ that splits the tree such
  that the part containing the root of $T$ and $v$ has size at most
  $\lceil n/2\rceil$ and such that all other components have size at
  most $\lfloor n/2\rfloor$. For the embedding take the splitting
  vertex $v\in T$ given by \Cref{lemsplit} which will be
  embedded to the marked vertex of the copy of
  $ S_{\lceil\frac{n}{2}\rceil} ' $ (cf.\
  \Cref{fig:basicrecursion}). Consider the three components of $T$
  after splitting. By the induction hypothesis there is an embedding
  function to embed the child components (components below the
  splitting vertex $v$) into the two copies of
  $S_{\lfloor \frac{n}{2}\rfloor} $. Then we take the component with
  the root and the marked vertex to get a tree with a single marked
  leaf of size at most $\lceil\frac{n}{2}\rceil$ and the induction
  hypothesis again provides with an embedding function of this
  component into $ S_{\lceil\frac{n}{2}\rceil} ' $. The embedding is
  depicted in \Cref{fig:Sn_embedding}.

  \begin{figure}[t]
    \centering
    \includegraphics[width=0.8\textwidth]{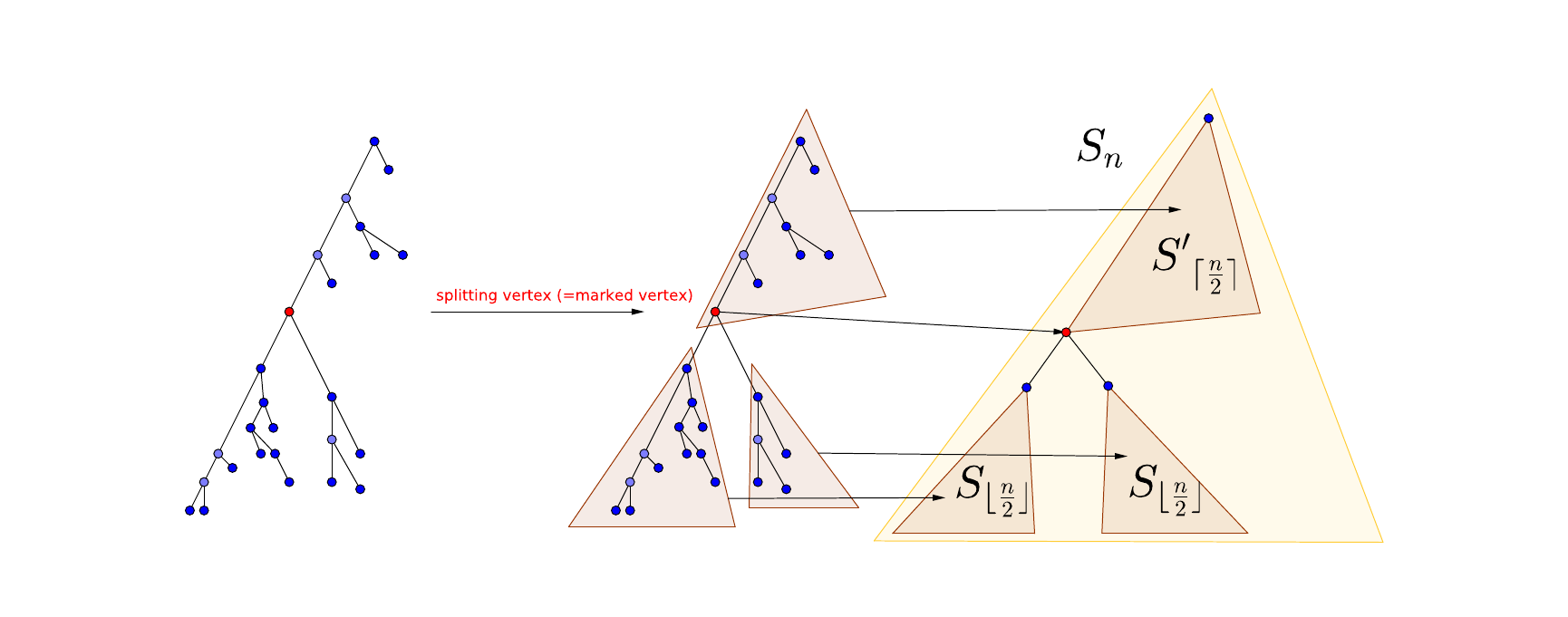}
    \caption{Embedding of an $n$-vertex binary tree $T$ into $S_n$.}
    \label{fig:Sn_embedding}
  \end{figure}

  To see that the NCA-function for any two vetices $u_1$ and $u_2$ and
  the described embedding function $\varphi_T$ commute, we need that
  $\nca_{S_n}(\varphi_T(u_1), \varphi_T(u_2)) =
  \varphi_T(\nca_{T}(u_1,u_2)) $. This can be verified through the
  following case analysis considering the three components after the
  splitting process.  If
  $u_1$ and $u_2$ are in the same component, then we have embedded them
  into the same subtree and by the induction hypothesis, we have
  $\varphi_T(\nca_T(u_1,u_2))=\nca_{S_n}(\varphi_T(u_1),\varphi_T(u_2))$.
  If $u_1$ and $u_2$ are in different child components, we have
  $\nca_T(u_1,u_2)=v$ and the embedding is therefore also correct
  because is embedded to the vertex marked in red in
  \Cref{fig:Sn_embedding}. Finally, if $u_1$ is from the component
  containing the root and vertex $v$ and $u_2$ is a vertex from
  a child component, we have $\nca_T(u_1,u_2)=\nca_T(u_1,v)$ and
  similarly  $\nca_{S_n}(\varphi_T(u_1),\varphi_T(u_2))=\nca_{S_n}(\varphi_T(u_1),\varphi_T(v))$
  and the embedding is therefore again correct by the induction
  hypothesis (applied to the partial embedding into the subtree
  $S_{\lceil n/2\rceil}'$).

  For the family of $n$-vertex binary trees with a marked leaf, the
  embedding into the recursively constructed tree $S_n'$ (cf.\
  \Cref{fig:basicrecursion}) works in similar way. Let $T'$ be a
  binary tree of size at most $n$ and with a marked leaf. We apply
  \Cref{lemsplitmark} with parameter $\lambda=1/2$ to $T'$ to obtain a
  vertex $v\in V(T')$ that splits $T'$ into a) a subtree of size at
  most $\lceil n/2\rceil$ that contains the root of $T'$ and that
  contains $v$ as a leaf vertex, b) a subtree size at most
  $\lfloor n/2\rfloor$ that is rooted at a child of $v$ and contains
  the marked leaf of $T'$, and c) a (possibly empty) subtree of size
  at most $n-1$ rooted at a child of $v$. The tree $T'$ is embedded
  into $S_n'$ by embedding vertex $v$ to the center red node
  separating the three recursive subtrees in
  \Cref{fig:basicrecursion}. The three subtrees resulting after
  splitting $T'$ are embedded into the three recursively constructed
  subtrees $S_{\lceil n/2\rceil}'$, $S_{\lfloor n/2\rfloor}'$, and
  $S_{n-1}$ in the natural way. The proof that the embedding is
  correct is done in the same way as for the embedding of $T$ into
  $S_n$. The details of the embedding are illustrated in
  \Cref{fig:Snp_embedding}.

  \begin{figure}[t]
    \centering
    \includegraphics[width=0.8\textwidth]{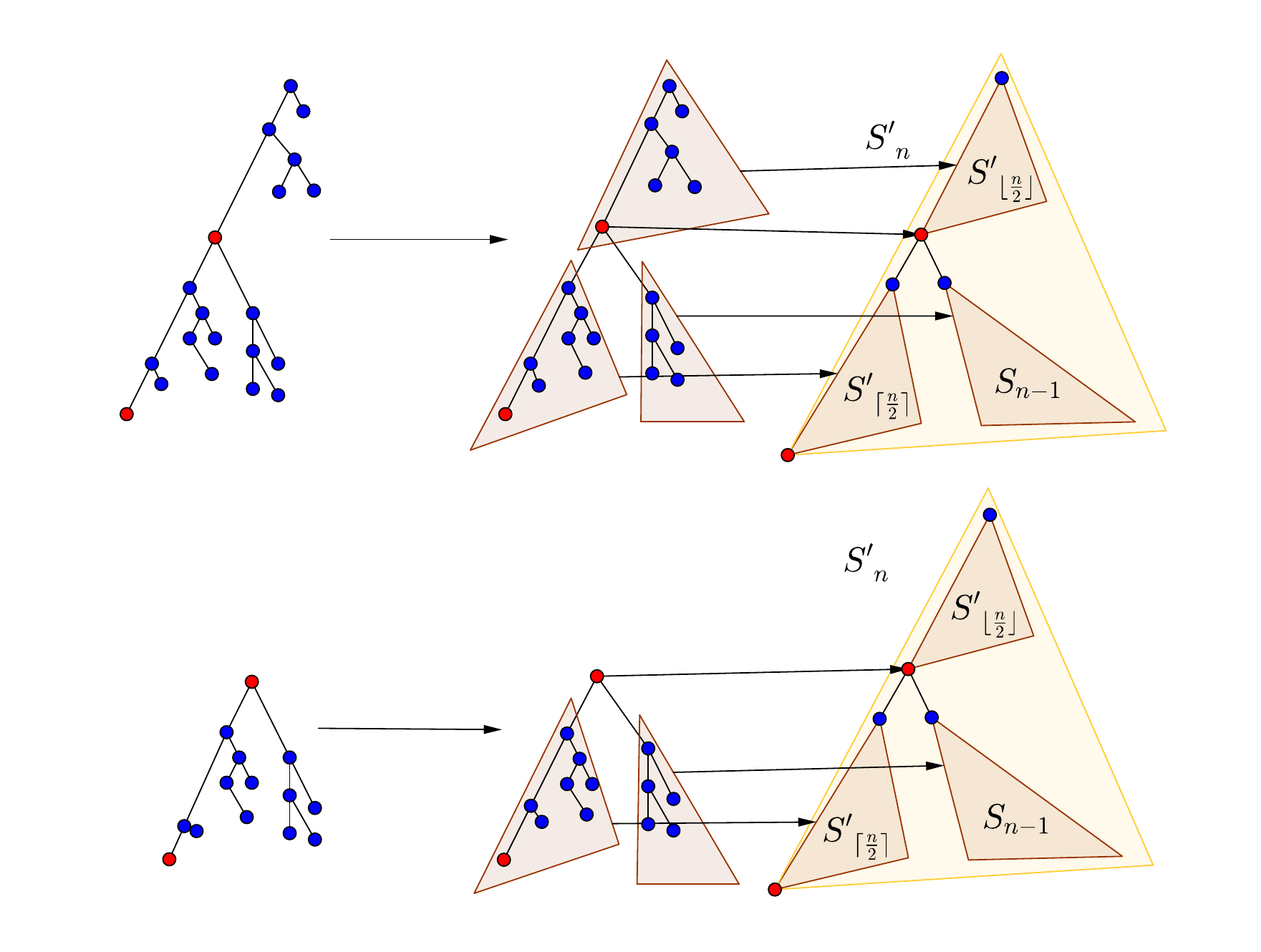}
    \caption{Embedding of an $n$-vertex binary tree $T'$ with a marked
      leaf into $S_n'$.}
    \label{fig:Snp_embedding}
  \end{figure}
\end{proof}

\section{General Universal Tree Construction}
\label{sec:universal}

The proof of the basic construction was handled in detail. We now adjust the construction to improve on the exponent in the size of $S_n$ and to deal with general rooted trees. Throughout the section, we omit floor and ceiling functions. They do not change the calculations significantly, but hinder the readability of the proof.

\begin{proof}[{\rm\bf Proof of \Cref{thm:universal}}]
  As suggested in \Cref{lemsplit} and \Cref{lemsplitmark}, the
  adjustment of the basic construction can be made by choosing $\lambda \neq 1/2$ for the
  size of the splitted components.

  We start with the binary tree case. We apply the same induction as
  in the proof of \Cref{thm:reduced}. In the general case, we prove that
  $|S_k| \le k^\beta$ and $|S_k'| \le c \cdot k^\beta$
  $ \forall ~1\le k < n$ and some constants $c$ and $\beta$ that will
  be determined later.

  \begin{figure}[t]
    \centering
    \includegraphics[width=0.8\textwidth]{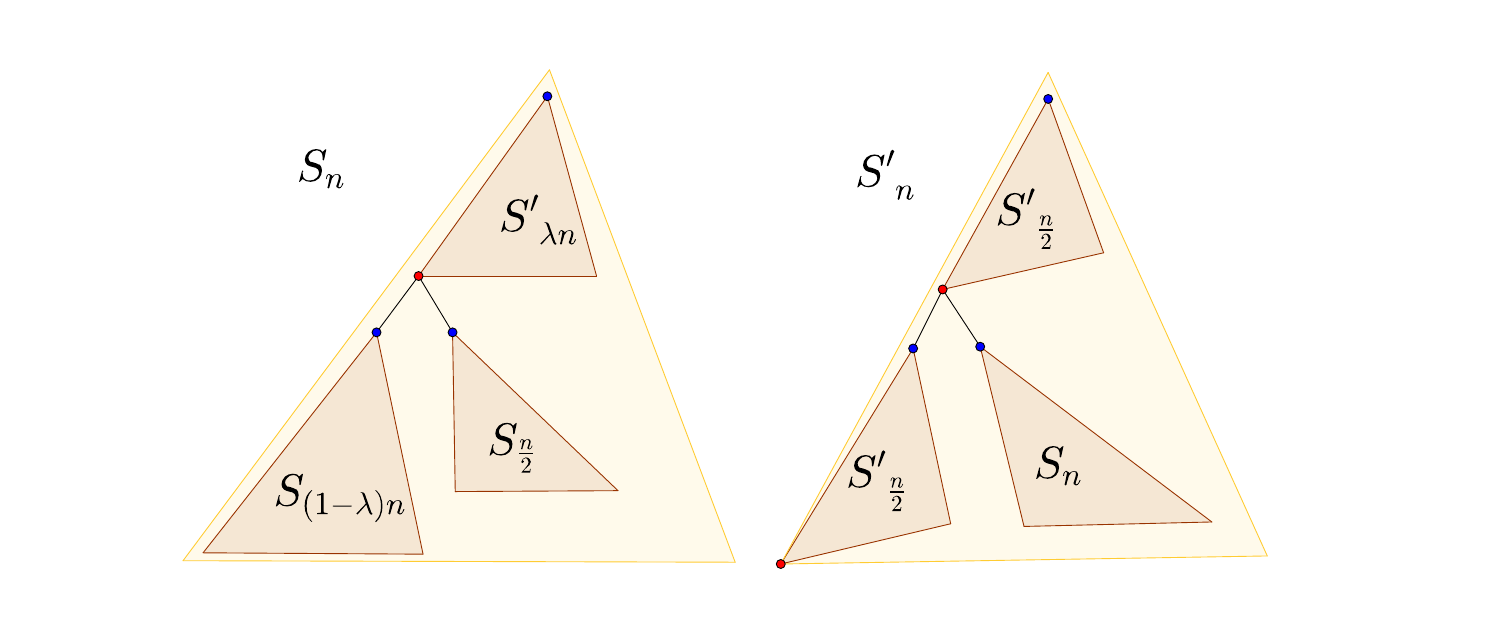}
    \caption{General recursive construction of an NCA-universal tree for $n$-vertex rooted binary trees.}
    \label{fig:Sn_lambdaembedding}
  \end{figure}

  For the construction of $S_n$ we take $ S_{\lambda n }' $ and attach
  copies of $S_{ (1-\lambda )n}$ and $ S_{{n}/{2}} $ to the marked
  vertex for some $\lambda \in (0,{1}/{2}]$. The construction of
  $S_n'$ remains the same as in \Cref{sec:simpleuniversal}. For an
  illustration, see \Cref{fig:Sn_lambdaembedding}. Note that although
  in \Cref{lemsplit} the child components can be of size
  $(1- \lambda) \cdot n$, the two components together can have size at
  most $n$, so the smaller of the components is always of size at most
  $ {n}/{2} $. We obtain
  $$ \begin{array}{llll}
       |S_n| &\le |S_{\lambda n }'| + |S_{ (1-\lambda )n}| +   |S_{\frac{n}{2}}  |  &\le c    (\lambda n)^\beta + {(  (1-\lambda ) n)}^\beta + { \left( \frac{n}{2} \right) }^\beta  &\stackrel{!}{\le}   n^\beta\\
       |S_n'| &\le |S_{ \frac{n}{2}}'| + |S_{\frac{n}{2}}'| +  |S_{n} |  &\le c{ \left( \frac{n}{2} \right) }^\beta + c { \left( \frac{n}{2} \right) }^\beta + n^\beta   &\stackrel{!}{\le} c n^\beta.
     \end{array}$$ 
     We would like to choose $\beta$ as small as possible. Note that with the second inequality, we obtain
$$ c{ \left( \frac{n}{2} \right) }^\beta + c { \left( \frac{n}{2} \right) }^\beta + n^\beta   \le c n^\beta $$
and we thus get that $c \ge \frac{1}{1-2^{\beta-1}} $.

In our construction we are allowed to freely choose $\lambda \in (0,{1}/{2}]$. We can thus choose $\beta$ and $\lambda$ such that $\beta$ is minimized and following inequality is still satisfied:
$$ { \ (1-\lambda )n}^\beta + c    \lambda n  ^\beta +\cdot \frac{n}{2}^\beta   \le  n^\beta.$$
To achieve this, we choose $\lambda = 0.296149...$ and the corresponding $\beta \le 1.89311...$. This proves the claim of \Cref{thm:universal} about binary rooted trees.

\begin{figure}[ht]
\centering
\includegraphics[width=0.8\textwidth]{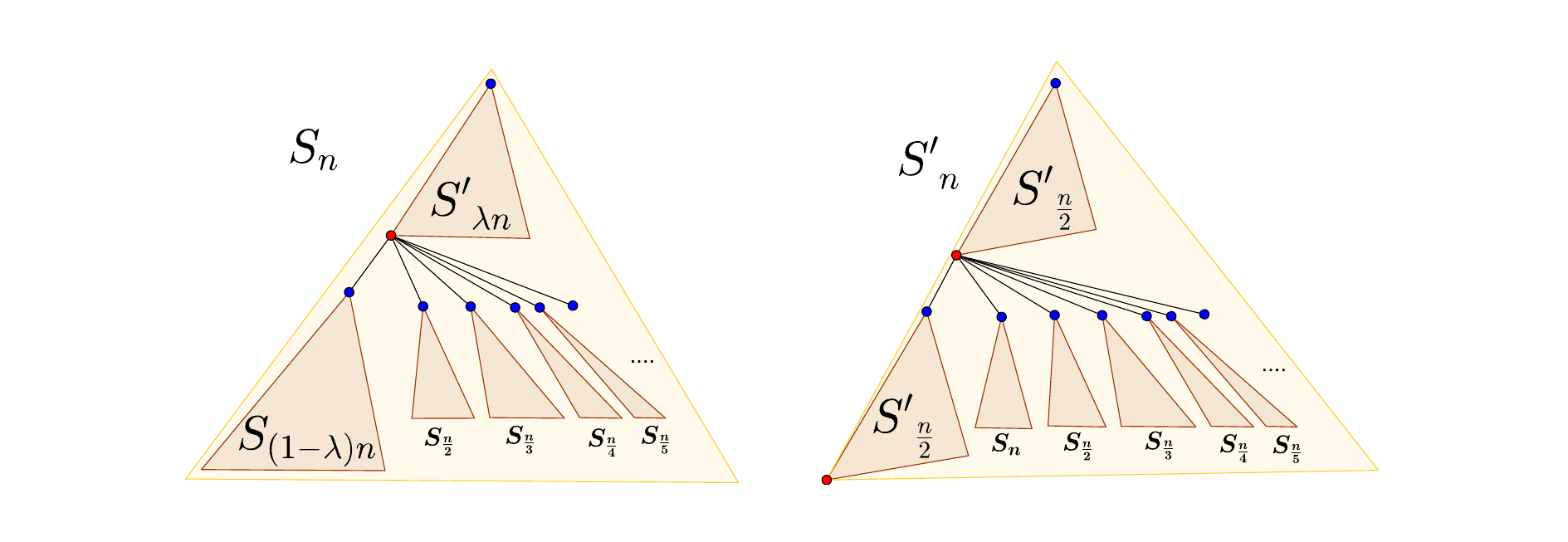}
    \caption{Construction of an NCA-univeral trees $S_n$ for general rooted trees.}
    \label{fig:recursionSnoptgen}
\end{figure}

In a general tree, a vertex can have many children. Therefore we adjust the construction to deal with this fact as shown in \Cref{fig:recursionSnoptgen}. Take any $\lambda \in (0,{1}/{2}]$. Let $S_n$ be composed of a copy of $S_{\lambda n}'$ and attached to the marked vertex of that tree copies of $S_{(1-\lambda)n}$, $S_{{n}/{2}}$, $S_{{n}/{3}}$, $S_{{n}/{4}}$, etc., up to $S_{1}$. Similarly, let $S_n'$ be composed of a copy of $S_{{n}/{2}}'$ and attached to the marked vertex of that tree copies of $S_{{n}/{2}}'$, $S_{n}$, $S_{{n}/{2}}$, $S_{{n}/{3}}$, $S_{{n}/{4}}$, etc., up to $S_{1}$. 
 
For the embedding, note that we can sort the child components by size.  \Cref{lemsplit} states that any child component is of size at most $(1-\lambda) \cdot n$. In addition, the total size of all components cannot add up to more than the entire tree of size $n$. This implies that after ordering the components by size, the $i^{\mathit{th}}$ child component without a marked vertex is of size at most ${n}/{i}$.

With the induction hypothesis that $|S_k| \le k^\beta$ and $|S_k| \le c \cdot k^\beta$, the recursion gives
$$ \begin{array}{llll}
|S_n| &\le |S_{ (1-\lambda )n}| + |S_{\lambda n }'| +  \sum_{i=2}^n |S_{\frac{n}{i}}  |  &\le {(\ (1-\lambda )n)}^\beta + c   (\lambda n) ^\beta +( \zeta(\beta) -1)  {n}^\beta &\stackrel{!}{\le}  n^\beta \\
|S_n'|& \le |S_{ \frac{n}{2}}'| + |S_{\frac{n}{2}}'| + \sum_{i=1}^n |S_{\frac{n}{i}}  |  &\le c{ \left( \frac{n}{2} \right) }^\beta + c { \left( \frac{n}{2} \right) }^\beta + \zeta(\beta) {n}^\beta  & \stackrel{!}{\le} c n^\beta,
\end{array},$$ 
where $\zeta(\beta)$ is the Riemann zeta function ($\zeta(\beta) =  \sum_{i\ge1} {i^{-\beta}}$ ). Again we can deduce from the second inequality that $c \ge \frac{\zeta(\beta) }{1-2^{\beta-1}} $. By using $\lambda = 0.341395...$, we get that $\beta \le 2.31757...$ .

In both constructions, the fact that the NCA function and
the embedding function commute follows in the same way as in the proof
of \Cref{thm:reduced}.  This concludes the proof of
\Cref{thm:universal}.
\end{proof}

\section{Implementation of the NCA-Labeling Scheme}
\label{sec:impl}

\begin{theorem}
The labeling schemes described in \Cref{sec:universal} can be constructed efficiently. Further, given two labels, the label of the nearest common ancestor can be determined in $O(\log^2n)$ time  (in $O(\log n)$ time in the binary tree case).
\end{theorem}
\begin{proof}[Proof Sketch] For $R \in \mathbb{R}$ we write $[R] := \{x: 1 \le x \le R\}$.
In order to assign labels to the vertices of $S_n$ we proceed as
follows. Set $s(n) =  n^\beta$, and  recall that $|S_n| \le s(n)$ and
$|S'_n| \le c s(n)$. Moreover, as depicted in \Cref{fig:recursionSnoptgen} the tree
$S_n$ is composed out of the $n+1$ trees
\begin{equation}
\begin{split}
T_0 & = S'_{\lambda n} \\
T_1 & = S_{(1-\lambda)n} \\
T_\ell & = S_{n/\ell}, \text{ where } 2 \le \ell \le n.
\end{split}
\end{equation}
Define the corresponding counting sequence
\begin{equation}
\begin{split}
t_{-1} & = 0 \\
t_0 & = c\, s(\lambda n) \\
t_1 & = t_0 + s((1-\lambda)n) \\
t_\ell & = t_{\ell-1} + s(n/\ell), \text{ where } 2 \le \ell \le n.
\end{split}
\end{equation}
For $S'_n$ we proceed similarly. As depicted in \Cref{fig:recursionSnoptgen} $S'_n$
is composed out of the $n+2$ trees
\begin{equation}
\begin{split}
T'_{-1} & = S'_{n/2} \\
T'_0 & = S'_{n/2} \\
T'_\ell & = S_{n/\ell}, \text{ where } 1 \le \ell \le n.
\end{split}
\end{equation}
and the corresponding counting sequence is given by
\begin{equation}
\begin{split}
t'_{-2} & = 0 \\
t'_{-1} & = cs(n/2) \\
t'_0 & = t_{-1} + cs(n/2) \\
t'_\ell & = t_{\ell-1} + s(n/\ell), \text{ where } 1 \le \ell \le n.
\end{split}
\end{equation}
Given these sequences, in order to assign labels to the vertices in
$S_n$ we assign to the vertices of $T_i, 0 \le i \le n$ the labels in
$[t_i] \setminus [t_{i-1}]$. The assignment is performed recursively, in
the sense that as soon the labels in $T_i, 0 \le i \le n$  are assigned,
they are translated by an additive $\lceil t_{i-1} \rceil$, so that they
all lie (with room to spare) in the required set $[t_i] \setminus
[t_{i-1}]$. The assignment is performed analogously for $S'_n$, where
we use the corresponding counting sequence instead.

Given the label of a vertex in $S_n$, its location in the tree can be
found with this preprocessing in $O(\log^2 n)$ time. Indeed, in every step
we have to decide in which of the at most n+2 subtrees we have to branch
to; however, this can be decided with binary search on the sequences
$(t_i)_{0 \le i \le n}$ or $(t'_i)_{-1 \le i \le n}$. As the depth of the recursive construction of $S_n$ is
$O(\log n)$, the claim follows. 
\end{proof}

\section{Conclusion}
\label{sec:concl}

We introduced NCA-universal trees and gave simple recursive
constructions of such trees that in particular lead to improved
NCA-labeling schemes for rooted trees. The paper leaves several
interesting open questions. The current upper bound of $2.318\log n$
bits per label is still quite far from the $1.008\log n$-bit lower
bound proven in \cite{Alstrup:2014:NLS:2634074.2634146} and it remains
an intriguing open problem to close this gap. In addition, given that
NCA-universal trees provide an intuitive way to argue about
NCA-labeling schemes, it is natural to ask whether the approach can
lead to optimal NCA-labeling schemes or whether every NCA-labeling
scheme for a given tree family can be turned into an equivalent one
that can be characterized by an NCA-universal tree for the tree
family. The following observation shows that NCA-universal trees are
equivalent to a certain well-structured class of NCA-labeling schemes.

We call an NCA-labeling scheme \textit{consistent} if any three labels can occur together in some tree. More formally, an NCA-labeling scheme is called consistent if it satisfies the following three properties for any 3 possible labels $x$, $y$, and $z$. In the following, $g$ is the decoder function.
\begin{enumerate}[(I)]
\item If $g(x,y)=z$, then $g(x,z)=z$ and $g(y,z)=z$\\ (i.e., 
  if $z$ is the NCA of $x$ and $y$, then $z$ is an ancestor of $x$ and $y$)
\item If $g(x,y)=y$ and $g(y,z)=z$, then $g(x,z)=z$\\ (i.e., 
  if $y$ is an ancestor of $x$ and $z$  an ancestor of $y$, then also $z$ is an ancestor of $x$)
\item If $g(x,y)=y$ and $g(x,z)=z$, then $g(y,z)\in\set{y,z}$\\
  (i.e., if $y$ and $z$ are ancestors of $x$, then $z$ is an ancestor
  of $y$ or $y$ is an ancestor of $z$)
\end{enumerate}

\begin{theorem}
  Every NCA-universal tree $S$ for a given family $\mathcal{T}$ of trees leads to a consistent NCA-labeling scheme for $\mathcal{T}$ with labels of size $\lceil\log|S|\rceil$. Conversely, every consistent NCA-labeling scheme for $\mathcal{T}$ and with $\ell$-bit labels induces an NCA-universal tree of size $2^{\ell}$ for $\mathcal{T}$.
\end{theorem}
\begin{proof}
  The first claim of the lemma is immediate because $S$ is a tree and therefore any three vertices of $S$ (i.e., any three labels) are consistent.

  For the second claim, define a directed graph $G=(V,E)$ as follows. The vertex set $V$ of $G$ is the set of labels of the given NCA-labeling scheme. Assume that $g$ is the decoder function of the labeling scheme.  We add a directed edge from $u\in V$ to $v\in V$ if $g(u,v)=v$ and there is no vertex $w$ such that $g(u,w)=w$ and $g(w,v)=v$ (i.e., if $v$ is the parent of $u$). We claim that $G$ is a rooted tree.

  First observe that $G$ is acyclic. Otherwise, by using Property (II) several times, we can find three vertices $u$, $v$, and $w$ such that $g(u,v)=v$, $g(v,w)=w$, and $g(w,u)=u$. However from Property (II) we then also have $g(u,w)=w$, a contradiction.

  Second, we show that the out-degree of each vertex of $G$ is at most $1$. For contradiction, assume that there exists a vertex $u$ that has out-going edges to $v$ and $w$. We then have $g(u,v)=v$ and $g(u,w)=w$ and by Property (III) of consistent labeling schemes, we thus also have $g(v,w)=w$ or $g(w,v)=v$. Thus, one of the two edges $(u,v)$ and $(u,w)$ cannot be in $G$.

  Finally, we show that there can be at most one vertex with out-degree $0$. For the sake of contradiction assume that $u$ and $v$ both have out-degree $0$ and let $g(u,v)=w$. By Property (I), we then also have $g(u,w)=w$ and $g(v,w)=w$. If $w\neq u$, this implies that $u$ has out-degree at least $1$ and if $w\neq v$, it implies that $v$ has out-degree at least $1$.

  Hence, $G$ is a rooted tree on the set of labels of the labeling scheme. Because the ancestry relationship of $G$ is consistent with the labeling scheme, $G$ is an NCA-universal tree for the family $\mathcal{T}$.
\end{proof}

\bibliographystyle{abbrv}
\bibliography{NCA}

\begin{thebibliography}{10}

\bibitem{Abiteboul:2001:CLS:365411.365529}
S.~Abiteboul, H.~Kaplan, and T.~Milo.
\newblock Compact labeling schemes for ancestor queries.
\newblock In {\em Proceedings of the Twelfth Annual ACM-SIAM Symposium on
  Discrete Algorithms}, SODA '01, pages 547--556, Philadelphia, PA, USA, 2001.
  Society for Industrial and Applied Mathematics.

\bibitem{Alstrup:2002:NCA:564870.564914}
S.~Alstrup, C.~Gavoille, H.~Kaplan, and T.~Rauhe.
\newblock Nearest common ancestors: A survey and a new distributed algorithm.
\newblock In {\em Proceedings of the Fourteenth Annual ACM Symposium on
  Parallel Algorithms and Architectures}, SPAA '02, pages 258--264, New York,
  NY, USA, 2002. ACM.

\bibitem{Alstrup:2014:NLS:2634074.2634146}
S.~Alstrup, E.~B. Halvorsen, and K.~G. Larsen.
\newblock Near-optimal labeling schemes for nearest common ancestors.
\newblock In {\em Proceedings of the Twenty-fifth Annual ACM-SIAM Symposium on
  Discrete Algorithms}, SODA '14, pages 972--982, Philadelphia, PA, USA, 2014.
  Society for Industrial and Applied Mathematics.

\bibitem{bender2000lca}
M.~A. Bender and M.~Farach-Colton.
\newblock The lca problem revisited.
\newblock In {\em LATIN}, volume 1776, pages 88--94. Springer, 2000.

\bibitem{berkman1993recursive}
O.~Berkman and U.~Vishkin.
\newblock Recursive star-tree parallel data structure.
\newblock {\em SIAM Journal on Computing}, 22(2):221--242, 1993.

\bibitem{1053860}
M.~Breuer.
\newblock Coding the vertexes of a graph.
\newblock {\em IEEE Transactions on Information Theory}, 12(2):148--153, April
  1966.

\bibitem{BREUER1967583}
M.~A. Breuer and J.~Folkman.
\newblock An unexpected result in coding the vertices of a graph.
\newblock {\em Journal of Mathematical Analysis and Applications}, 20(3):583 --
  600, 1967.

\bibitem{Fraigniaud:2010:CAL:1873601.1873639}
P.~Fraigniaud and A.~Korman.
\newblock Compact ancestry labeling schemes for xml trees.
\newblock In {\em Proceedings of the Twenty-first Annual ACM-SIAM Symposium on
  Discrete Algorithms}, SODA '10, pages 458--466, Philadelphia, PA, USA, 2010.
  Society for Industrial and Applied Mathematics.

\bibitem{Gabow:1984:SRT:800057.808675}
H.~N. Gabow, J.~L. Bentley, and R.~E. Tarjan.
\newblock Scaling and related techniques for geometry problems.
\newblock In {\em Proceedings of the Sixteenth Annual ACM Symposium on Theory
  of Computing}, STOC '84, pages 135--143, New York, NY, USA, 1984. ACM.

\bibitem{Gavoille2003}
C.~Gavoille and D.~Peleg.
\newblock Compact and localized distributed data structures.
\newblock {\em Distributed Computing}, 16(2):111--120, Sep 2003.

\bibitem{1707.06011}
P.~Gawrychowski and J.~Łopuszański.
\newblock Better labeling schemes for nearest common ancestors through
  minor-universal trees.
\newblock {\em arXiv:1707.06011}, 2017.

\bibitem{harel1980linear}
D.~Harel.
\newblock A linear time algorithm for the lowest common ancestors problem.
\newblock In {\em Foundations of Computer Science, 1980., 21st Annual Symposium
  on}, pages 308--319. IEEE, 1980.

\bibitem{hareldoi:10.1137/0213024}
D.~Harel and R.~E. Tarjan.
\newblock Fast algorithms for finding nearest common ancestors.
\newblock {\em SIAM Journal on Computing}, 13(2):338--355, 1984.

\bibitem{doi:10.1137/0405049}
S.~Kannan, M.~Naor, and S.~Rudich.
\newblock Implicat representation of graphs.
\newblock {\em SIAM Journal on Discrete Mathematics}, 5(4):596--603, 1992.

\bibitem{Peleg1999}
D.~Peleg.
\newblock Proximity-preserving labeling schemes and their applications.
\newblock In {\em Proc.\ 25th Int. Workshop on Graph-Theoretic Concepts in
  Comp.\ Sc.}, pages 30--41, 1999.

\bibitem{powell1990further}
P.~Powell.
\newblock further improved lca algorithm.
\newblock {\em Plant Genome Data and Information Center collection on
  computational molecular biology and genetics}, 1990.

\bibitem{Rado1964}
R.~Rado.
\newblock Universal graphs and universal functions.
\newblock {\em Acta Arithmetica}, 9(4):331--340, 0 1964.

\bibitem{doi:10.1137/0217079}
B.~Schieber and U.~Vishkin.
\newblock On finding lowest common ancestors: Simplification and
  parallelization.
\newblock {\em SIAM Journal on Computing}, 17(6):1253--1262, 1988.

\end{thebibliography}

\newpage

\appendix
\section{Appendix}
\subsection{Induction Basis}
\label{app:basis}
For n equals one through four it is simple to find a universal tree of size at most $n^2$. For example:

\begin{figure}[h]
\centering
\includegraphics[width=12cm]{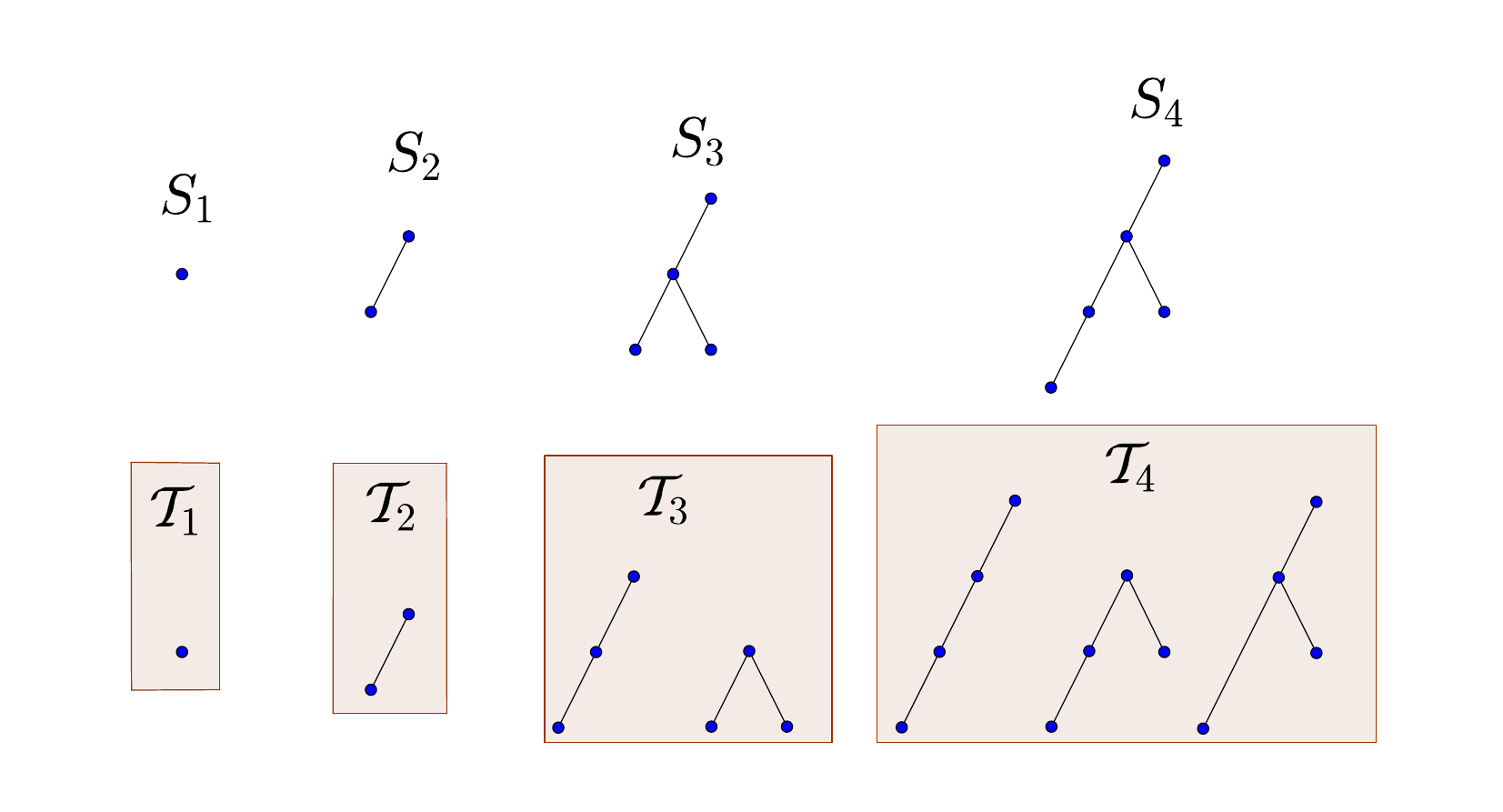}
\caption{The root is always the top most vertex but the root of $T \in \mathcal{B}_n$ does not have to be embedded to the root of $S_n$}
\end{figure}

$S_n'$ shall have size at most $2n^2$. We want for any n a large tree $S_n'$ with a marked leaf such that we can embed any tree on n vertices with a marked leaf with the nca-function commuting. Again as example and induction basis:

\begin{figure}[h]
\centering
\includegraphics[width=12cm]{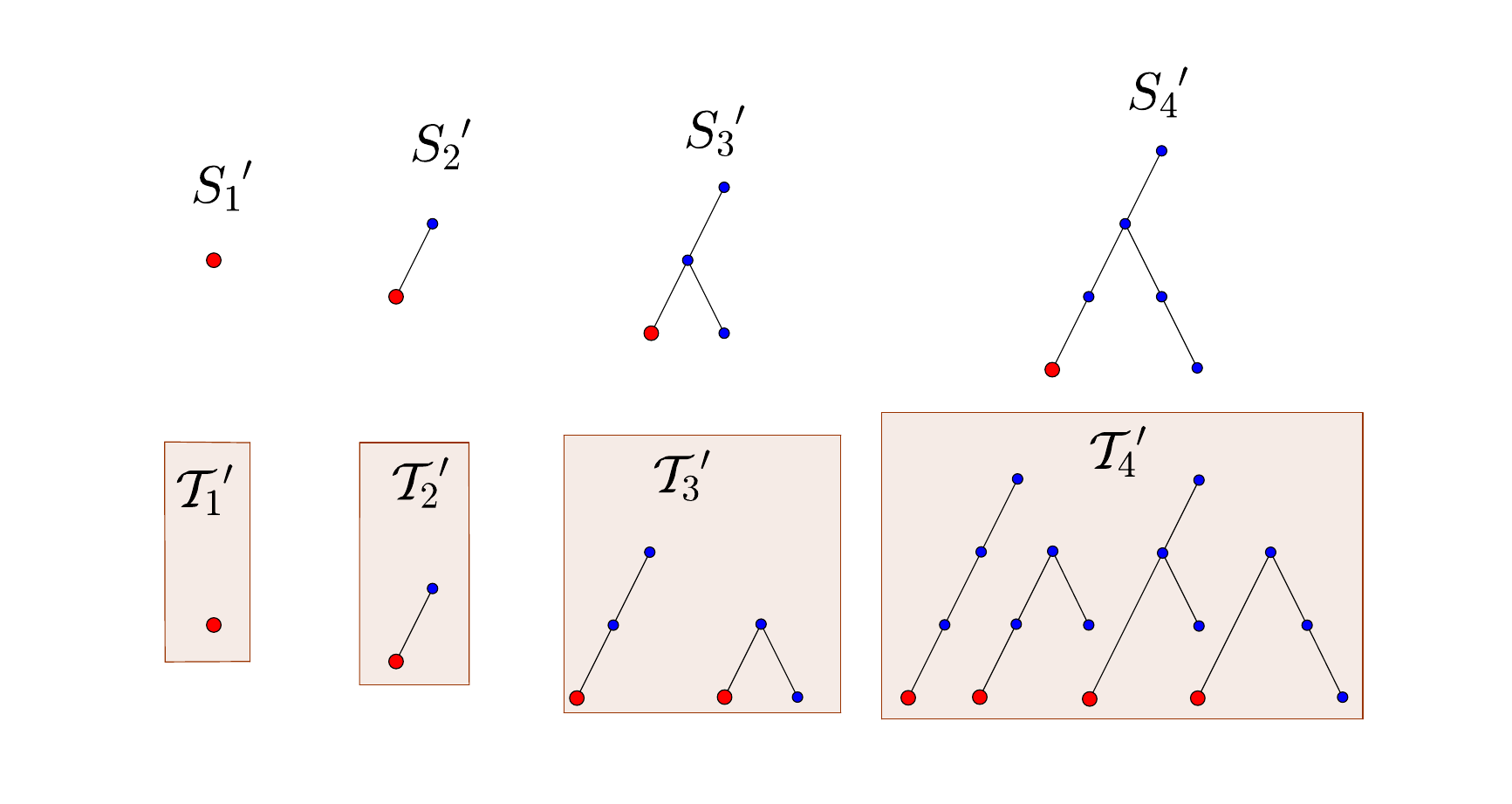}

\caption{  The larger red vertex is the marked leaf.}
\end{figure}

It is easy to check that $\forall T \in \mathcal{B}_l $ and $\forall T' \in \mathcal{B}_l' $ the NCA-query is equivalent to the NCA-query in $S_l$ or $S_l'$ $\quad   \quad   \forall l \in \{1,.. \ , 4\}$ resp.

\end{document}